\documentclass[12pt]{article}
\usepackage{graphicx}
\begin{document}

\centerline{\bf Simulation of stem cell survival in small crypts}

\bigskip
Dietrich Stauffer* and Eytan Domany

\bigskip
Physics of Complex Systems, Weizmann Institute of Science  

76100 Rehovot, Israel
\bigskip

*Visiting from Institute for Theoretical Physics, Cologne University,

D-50923 Cologne, Euroland

\bigskip
\centerline {Abstract}

{\small Monte Carlo simulations of the number of stem cells in human colon 
crypts allow for fluctuations which kill the population after sufficiently long 
times.}

\bigskip
Experiments and computer simulations \cite{yatabe,ro} have elucidated the 
fate of stem cells and of maturing cells within small crypts in the human colon.
The present note restricts itself on the simulation of the number of cells,
not on their genealogy.

Yatabe et al. \cite{yatabe,ro} assume that at each iteration (``day'') each stem
cell splits into two stem cells with probability $q$, into two maturing cells
with the same probability $q$, and into one stem cell plus one maturing cell 
with probability $p = 1 - 2q$. The special case $p=1, \; q=0$ is deterministic 
and not dealt with by us. For simplicity we assume $p = 0$, which means
$q = 1/2$ in the model studied by Yatabe et al. 

As soon as probabilities larger than zero and smaller than one are used, one
introduces fluctuations. Sometimes the fluctuations can become very large and in
the present model kill all stem cells \cite{redner}. From that moment on the 
crypt remains without stem cells, and after several iterations also the 
remaining maturing cells will have reached their life span, causing the whole
crypt to die. For a functioning biological system this death of the 
whole crypt should not happen too often within reasonable time. For
infinitely long times, every finite system living according to the above rules 
will finally die out; we are interested here in the survival for 365 iterations
(one ``year''). 

In addition to the above random splitting, we also allow for random deaths
happening with a probability of one percent per iteration for each cell,
independent of age and type. Moreover, maturing cells die when their life span
of three iterations is exceeded. In this case, the number of cells in a system
of initially 100 million cells decreases by one percent per iteration, and we 
balance this decrease by making the model slightly asymmetric: With probability 
$q_1 = 0.5 + b$ stem cells produce two stem cells each, and with probability 
$q_2=0.5-b$ they produce two maturing cells each, with  a small bias $b=0.01$. 
Now an initial population of 10 million stem cells produces within a few
iterations more than 100 million maturing cells, and from then on both numbers
stay nearly the same for 100 iterations (Fig.1).

However, Yatabe et al applied their model to crypts with numbers $N$ of stem
cells as small as $N = 4$. In that case, extinction is the rule and not the 
exception: In one million cases simulated, we found survival of the whole
crypt for 365 iterations in only 1, 5 and 12 percent for 4, 10 and 25 initial
stem cells. The double-logarithmic Fig.2 shows the distribution of extinction 
times, which for long times seems to follow a power law with an effective
exponent near --1.9 (straight line) \cite{redner}. 

We tried to improve the survival chances by replacing the above bias $b$ by a 
linear feedback: The probability $q_2 = 1-q_1$ for a stem cell to split into two
maturing cells was taken as $N(t)/2N_0$  where $N(t)$ is the actual number of
stem cells at iteration $t$ and $N_0$ is the desired and initial number of such
stem cells. Then for $N_0$ = 4 the situation got worse: None of the one million 
cases survived even to half a year; but for $N_0=10$ and 25 the survival chances
increased to 26 and 99.95 percent, respectively. The semilogarithmic Fig.3 
compares the distributions of extinction times in this feedback version
with that of the previous constant bias $b$, for $N_0 = 4$, 10 and 25. (10
million samples for $N_0 = 25$.)

In summary, it is difficult to keep only four stem cells surviving for a long 
time, but for 10 and 25 stem cells, the linear feedback offers a good to
excellent survival chance. All these fluctuation problems would vanish for
the deterministic case $p=1$. 

The German-Israeli Foundation supported this collaboration.

\bigskip
\centerline{\bf Appendix}

In this appendix we give some computational details of the programs 
available from stauffer@thp.uni-koeln.de:
stem2.f (random) and stem3.f (re\-storing linear bias) 

For the random choice without a bias depending on the actual number $N(t)$ of 
stem cells, in each iteration a first loop determines all the deaths due to
the 1 \% death probability and, for maturing cells, due to having
surpassed the maximum lifespan of three. $N_t$ is the total number of living
cells, stem cells plus maturing cells. For this purpose, going backward from 
$N_t$ to 1 in the index $i$ for the individual cells, in the case of death
individual $i$ is replaced by the last individual in the line, i.e. by the
survivor $i_{max}$ with the highest index. Replacing means to increase the
number of deaths by one and to transfer the current age of $i_{max}$ to $i$.
At the end of that loop, $N_t$ is diminished by the number of deaths.

Then a second loop from $i=1$ to $N_t$ simulates births in two steps. First, the
number of births is increased by one. Second, a random number determines 
whether a stem cell splits into two maturing cells or into two stem cells; the 
first choice is always taken if $i$ refers to a maturing cell. In the first 
case both new cells then have an age higher by one than the parent cell; in the
second case both stem cells keep their age at zero. At the end of this second 
loop, $N_t$ is increased by the number of births, and a third loop finds the
number of stem cells (defined as having zero age). Then the numbers of deaths 
and births are reset to zero, and the next iteration starts. 

For the biased version, with a probability of stem cells to divide into two
maturing cells increasing linearly with the number of stem cells, the first
(death) loop remains unchanged. Then the number $N_t$ of surviving cells and
the number $N(t)$ of surviving stem cells as well as the resulting probability 
$q_2 = N(t)/2N_0$ to split into two maturing cells are determined. With this
constant $q_2$ valid for all cells $i = 1,2, \dots, N_t$ at this iteration, 
the above birth loop (second loop for unbiased random version) produces new 
cells.

We also used a third version where, as in the Penna ageing model, deaths and
births are not separated into two consecutive loops but combined \cite{book}.
The same bias, increasing linearly with $N(t)$ and constant for all cells
within one iteration, is used. Then survival becomes somewhat easier, Fig.4,
with 21 out of 10 million surviving one year at $N_0 = 4$, and 6 out of one
million dying at $N_0 = 25$. With this method one may also use sequential 
updating of the restoring linear bias, that means the bias is not fixed during
one whole iteration but fluctuates with the fluctuating stem cell number by
being updated after every single stem cell division. Then 1.5 percent of ten
million samples with $N_0 =4$ survive one year. This survival fraction increases
to 26 \% if the restoring bias is applied only for $N(t) < N_0$ and not for
$N(t) \geq N_0$.

\begin{figure}[hbt]
\begin{center}
\includegraphics[angle=-90,scale=0.5]{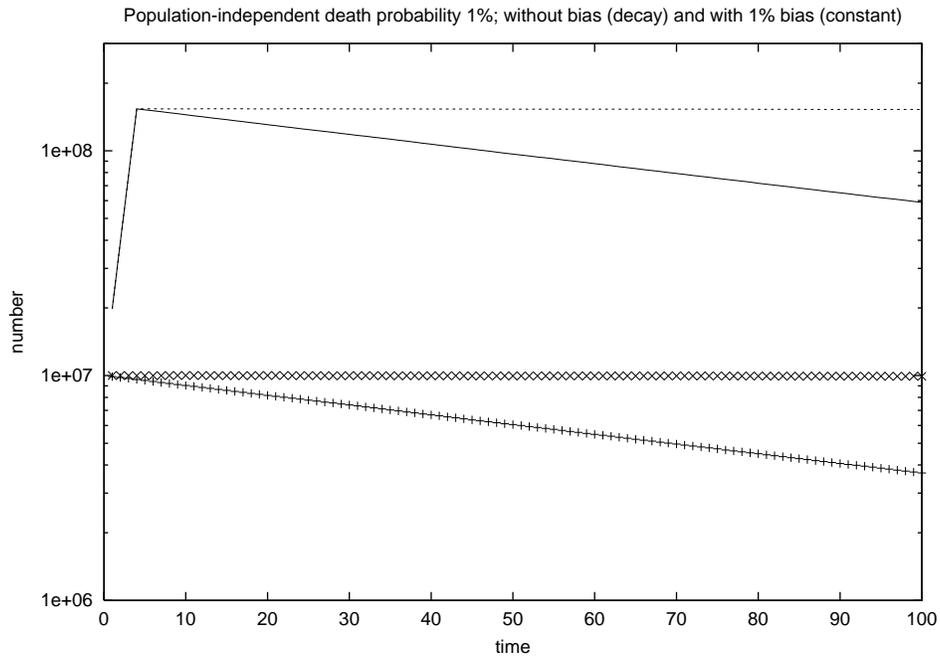}
\end{center}
\caption{Number of maturing cells (top) and stem cells (bottom) for large
populations.
}
\end{figure}

\begin{figure}[hbt]
\begin{center}
\includegraphics[angle=-90,scale=0.5]{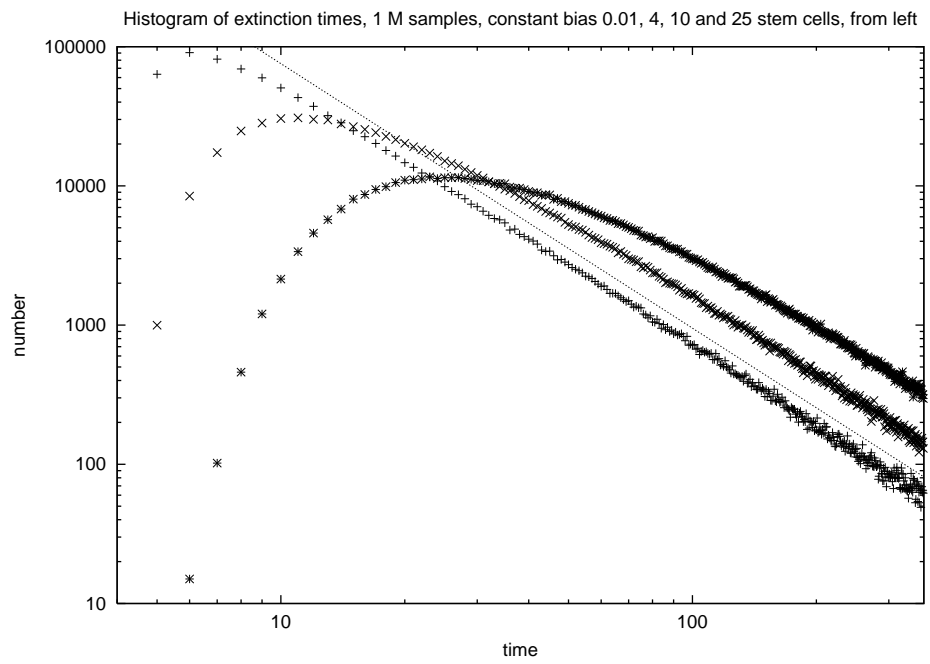}
\end{center}
\caption{Extinction times, increasing with increasing $N_0 = 4$, 10 and 25
(double-logarithmic plot using one million samples).}
\end{figure}

\begin{figure}[hbt]
\begin{center}
\includegraphics[angle=-90,scale=0.5]{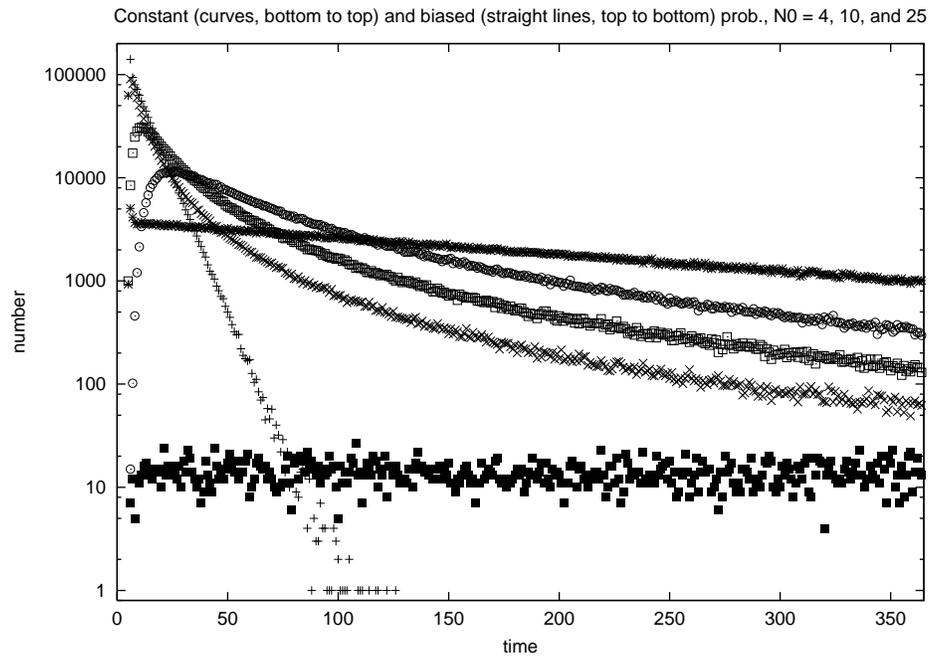}
\end{center}
\caption{
Semi-logarithmic plot of extinction times without (curves) and with 
(straight lines) linear restoring bias. For $N_0$ = 25 with bias, 10 million
instead of one million samples were simulated.}
\end{figure}

\begin{figure}[hbt]
\begin{center}
\includegraphics[angle=-90,scale=0.5]{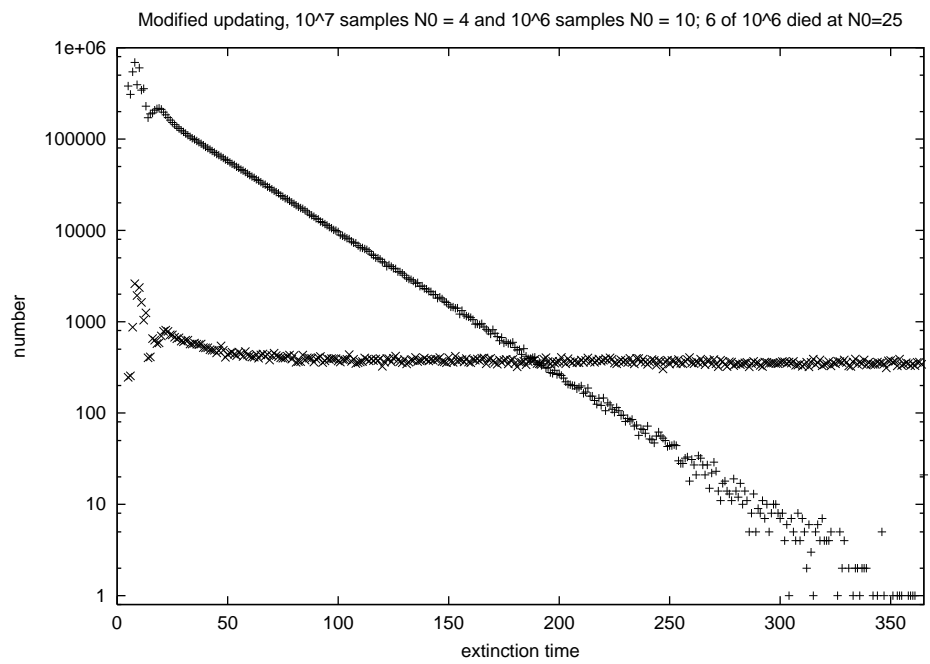}
\end{center}
\caption{Semi-logarithmic plot of extinction times in modified updating.
}
\end{figure}

\end{document}